\documentclass[aps,tightenlines,nofootinbib,prd]{revtex4}
\pdfoutput=1
\usepackage{amsmath,amscd,amssymb}
\usepackage{color,epsfig}
\usepackage{xfrac}
\usepackage{latexsym}
\usepackage{mathrsfs,bigints}
\usepackage{graphicx}
\usepackage{bbm}      

\newcommand{\imu}{{\rm i}}

\usepackage[dvipsnames]{xcolor}

\begin{document}

\title{Quantum Effects of Solitons in the Self-Dual Impurity Model}

\author{I. Takyi$^{a)}$, H. Weigel$^{b)}$}

\affiliation{$^{a)}$Mathematics Department, Kwame Nkrumah University of Science and 
Technology, Private Mail Bag, University Post Office, KNUST-Kumasi, Ghana\\
$^{b)}$Institute for Theoretical Physics, Physics Department, 
Stellenbosch University, Matieland 7602, South Africa}

\begin{abstract}
We compute the vacuum polarization energies (VPE) of solitons in a self-dual impurity model in which the
soliton profiles take the shape of a separated kink-antikink pair. Classically the soliton energies are 
invariant under the change of a continuous parameter that can be interpreted as the kink-antikink separation. 
This is not the case for the VPE so that quantum effects decide on the energetically most favorable 
separation. The considered configurations are classically stable so that its quantum fluctuations have 
only real frequency eigenvalues. Hence, in contrast to the kink-antikink configuration in the $\phi^4$ 
model, the VPE is well defined for any value of the separation and we gain insight into the quantum 
corrections to the kink-antikink potential.
\end{abstract}

\maketitle

\section{Introduction}
In this work we explore the one-loop quantum corrections to the energies of solitons in a self-dual impurity 
$\phi^{4}$ model in one-space and one-time dimension ($D=1+1$). Low dimensional soliton models are interesting
because they are role models for more complex systems in higher dimensions that have applications ranging from, 
among others, cosmology~\cite{Vilenkin:2000jqa} via condensed matter physics~\cite{Schollwock:2004aa,Nagasoa:2013} 
to hadron~\cite{Weigel:2008zz} and nuclear physics~\cite{Feist:2012ps}. An important feature of the model that we 
consider here is that its solitons saturate a Bogomolny-Prasad-Sommerfield (BPS) \cite{Bogomolny:1975de,Prasad:1975kr} 
energy bound. Hence the classical energy only depends on the boundary values of soliton profiles but not on the position 
of the impurity so that there is no static force between the impurity and the soliton~\cite{Adam:2018pvd,Adam:2018tnv}. 
As a further consequence these solitons are degenerate with respect to the variation of a continuous parameter that 
measures the distance between the position of the kink-type structure and the center of the impurity. Thus the most 
favorable solution is determined by the quantum corrections to the classical energy even though these corrections 
are small for a consistent choice of model parameters. Computing these corrections as a function of this variational 
parameter is the central objective of this work. For a sufficiently strong impurity we will find a local minimum of 
the quantum energy. However, for weak impurities, this energy favors infinitely far separated kink-antikink pairs, 
signaling an unstable soliton. This is similar to multi-field Shifman Voloshin soliton~\cite{Shifman:1997wg}
that has classically degenerate solitons which are unstable quantum mechanically~\cite{Weigel:2018jgq}.
Similarly, quantum corrections reveal instabilities in higher polynomial soliton 
models~\cite{Weigel:2016zbs,Takyi:2020tvl}. 

For moderate and large values of the separation parameter the static solutions resemble superpositions of a
kink-antikink pair in the renowned $\phi^{4}$ model~\cite{Campbell:1983xu,Anninos:1991un,Takyi:2016tnc}. In contrast 
to kink-antikink superpositions in the $\phi^{4}$ model the impurity causes such configurations to be stable 
classically so that the model provides a way around the fundamental problem of dealing with imaginary 
frequencies~\cite{Graham:1998kz,Lee:2016lhd} of the quantum fluctuations about a kink-antikink in the $\phi^{4}$ model. 

The computation of quantum corrections to soliton energies in one space dimension is by now a standard and straightforward
endeavor when utilizing the so-called spectral methods~\cite{Graham:2009zz}. The essential ingredients are the scattering 
data extracted from the quantum fluctuations about the potential that is induced by the soliton. This computation
is particularly simple when this potential is invariant under spatial reflection~\cite{Graham:2022rqk}.
All what is needed is the Jost function along the positive imaginary momentum axis. Since the quantum 
energy manifests itself through the shift (or polarization) of the zero point energies of the 
quantum fluctuations it is frequently called the vacuum polarization energy (VPE).

Following this introduction, we will describe the self-dual impurity model and its BPS solutions in 
section \ref{sec:models}.  Section \ref{sec:pot} contains the analysis of the potential for the quantum
fluctuations at moderate and large kink-antikink separation. In section \ref{sec:vpe} we will describe the spectral 
method that determines the VPE. We will present the numerical results for the VPE of BPS solutions in the self-dual 
impurity model in section \ref{sec:results}. In section \ref{sec:KAK} we estimate the leading quantum correction 
to the kink-antikink potential. We conclude and summarize in section \ref{sec:conclude}. 

\section{The Model}
\label{sec:models}

We consider the self-dual impurity model in $D=1+1$ whose full Lagrangian is given 
by~\cite{Adam:2019djg,Adam:2019prh}
\begin{equation}
L = \int dx\, \left[\frac{1}{2} \left(\frac{\partial \phi}{\partial t}\right)^2
-\frac{1}{2} \left(\frac{\partial \phi}{\partial x} + \sqrt{2} W 
+ \sqrt{2}\, \sigma W \right)^{2} \right] + \int dx\, \sqrt{2} W\frac{\partial \phi}{\partial x}\,,
\label{eq:Lag} \end{equation}
where $W=W(\phi)$ is the super-potential and $\sigma=\sigma(x)$ is the prescribed impurity. In the limit
$\sigma\,\to\,0$ the standard scalar model with a scalar potential of $U=W^{2}$ is obtained. Typically these potentials
generate spontaneous symmetry breaking with at least two degenerate vacua. The soliton(s) then assume either
of them at positive and negative spatial infinity. We have written this Lagrangian in terms of dimensionless 
variables and parameters, thereby omitting an overall factor. This factor does not affect the field equations 
but, through canonical quantization, it is the order parameter that counts the loops entering the quantum 
corrections and thus discriminates between classical and quantum contributions. In our case we will solely compare 
configurations which are classically degenerate at one loop order. (Actually all configurations have zero classical 
energy.) Hence our results will not be sensitive to this order parameter and we may safely omit it.\footnote{In general, 
this parameter should be chosen to make the leading quantum correction small; otherwise higher order effects will 
presumably be sizable.}

The topological lower bound of the classical energy is saturated when the following Bogomolny equation holds 
\begin{equation}
\frac{\partial \phi}{\partial x} + \sqrt{2} W + \sqrt{2} \sigma W = 0\,.
\label{eq:Bogequ}
\end{equation}
In this study we take the super-potential to be
\begin{equation}
W = \frac{1}{\sqrt{2}} \left(1-\phi^{2}\right)\,,
\end{equation} 
which is of the $\phi^{4}$ model type in the no impurity limit. In particular, we consider the 
non-localized impurity~\cite{Adam:2019djg,Adam:2019prh}
\begin{equation}
\sigma_{j}(x)=\frac{j}{2} \tanh (x)-1\,,
\end{equation}
where $j \geq 0$ is the parameter that measures the strength of the impurity.
The BPS solutions in this self-dual impurity model are
\begin{equation}
\phi(x)=-\frac{\cosh^{j}x-a}{\cosh^{j}x+a}\,.
\label{eq:BPS_soliton} \end{equation}
We may take either $a \in \left(-1, \infty\right)$ or
\begin{equation}
a = -1 + {\rm e}^{jx_{0}} \quad {\rm with}\quad x_{0} \in \mathbb{R}\,,
\end{equation}
as the variation parameter that parameterizes the shape of the BPS soliton profiles.
\begin{figure}
\centerline{
\epsfig{file=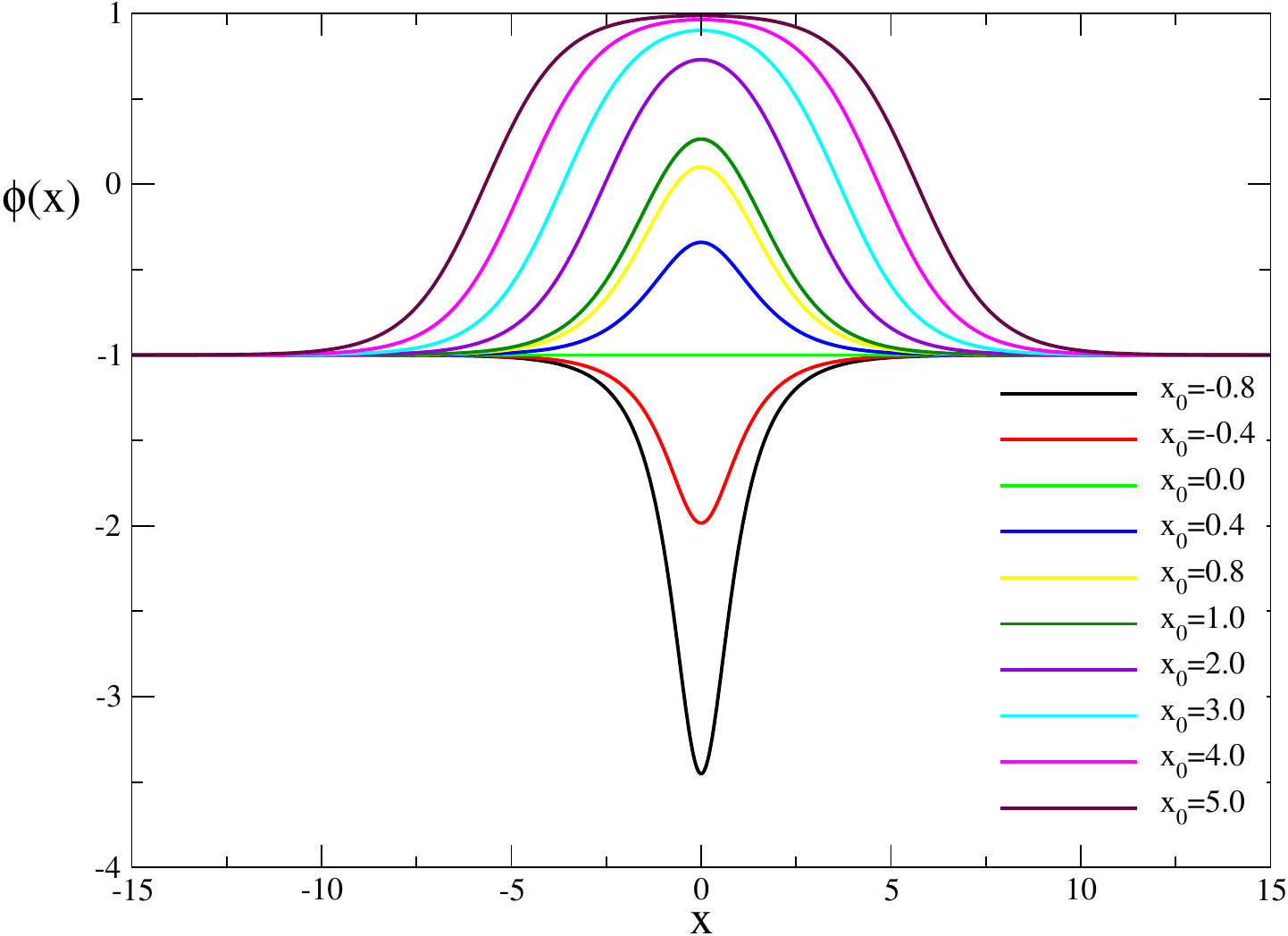,width=6cm,height=6cm}\hspace{0.2cm}
\epsfig{file=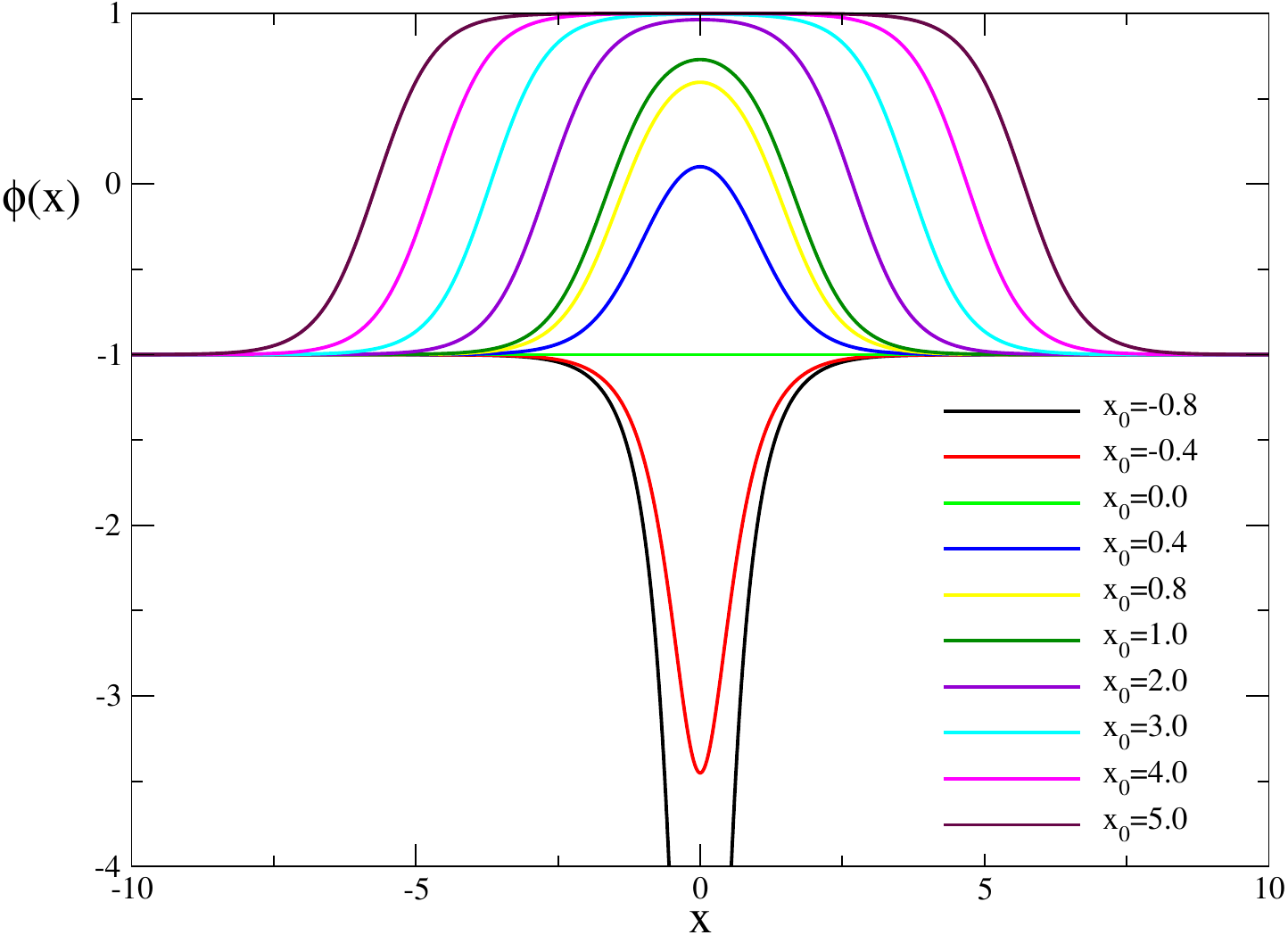,width=6cm,height=6cm}\hspace{0.2cm}
\epsfig{file=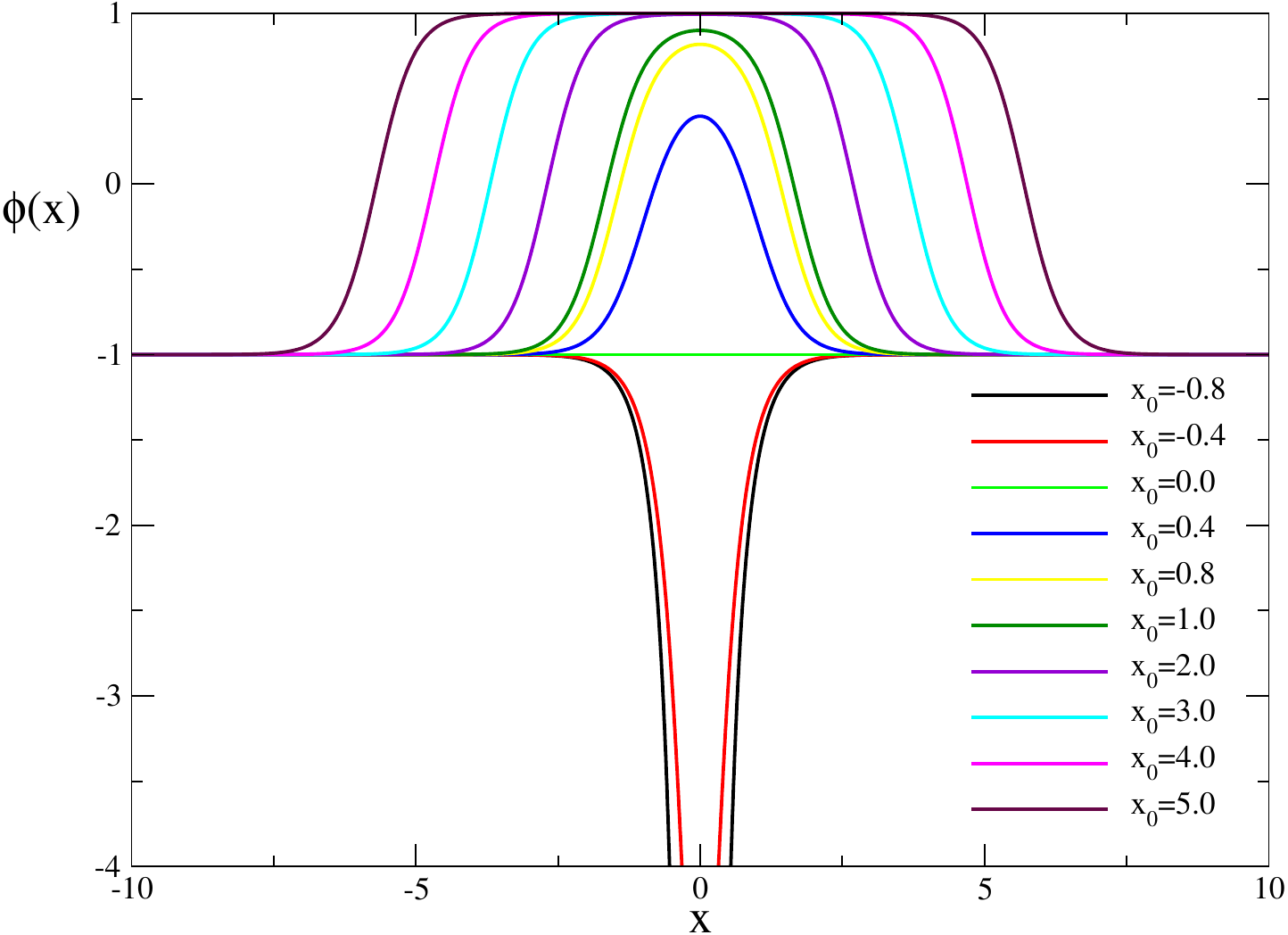,width=6cm,height=6cm}}
\caption{\label{fig:profile}(color online) From left to right: The profile of the BPS soliton in 
Eq. \eqref{eq:BPS_soliton} for various values of the inter-soliton distance ($x_{0}$) for $j=1$, $j=2$ 
and $j=3$.}
\end{figure}

For $x_{0}\gg1$ the BPS solutions represent infinitely widely separated kink-antikink pairs of the $\phi^{4}$ 
model as shown in Figure~\ref{fig:profile}. Also, as $x_{0}$ tends to $0$, the kink and antikink approach each 
other. For small impurities with $0<x_{0}\,{\scriptstyle\lesssim}\,1$, a BPS solution is similar to a kink-antikink 
pair in the $\phi^4$ model with small separation. However, for large impurities, the BPS solutions 
develop wells at the center. The variation parameter $x_{0}$ was interpreted as the distance between the BPS 
soliton and the impurity~\cite{Adam:2019prh} because of these particular features.

In general BPS solutions saturate the topological energy bound and the static energy thus only depends on the boundary values of the soliton. In our case this has the important consequence that this energy
is independent of the strength $j$ of the impurity. Hence there is no (static) force between the BPS soliton and 
the impurity. 

Obviously the super-potential vanishes at $\phi=\pm1$ which are the two possible vacua. At spatial infinity
the soliton approaches $\phi=-1$ which we thus call the primary vacuum. The equality between the profiles 
at positive and negative spatial infinity actually leads to a vanishing classical energy. Unless $x_0$ is very small, 
the soliton also occupies the secondary vacuum at $\phi=+1$ in a sizable region of space. This region grows with $x_0$.
The possibility of occupying such a secondary vacuum has been recently related to quantum destabilization of 
solitons \cite{Weigel:2018jgq}.

\begin{figure}
\centerline{
\epsfig{file=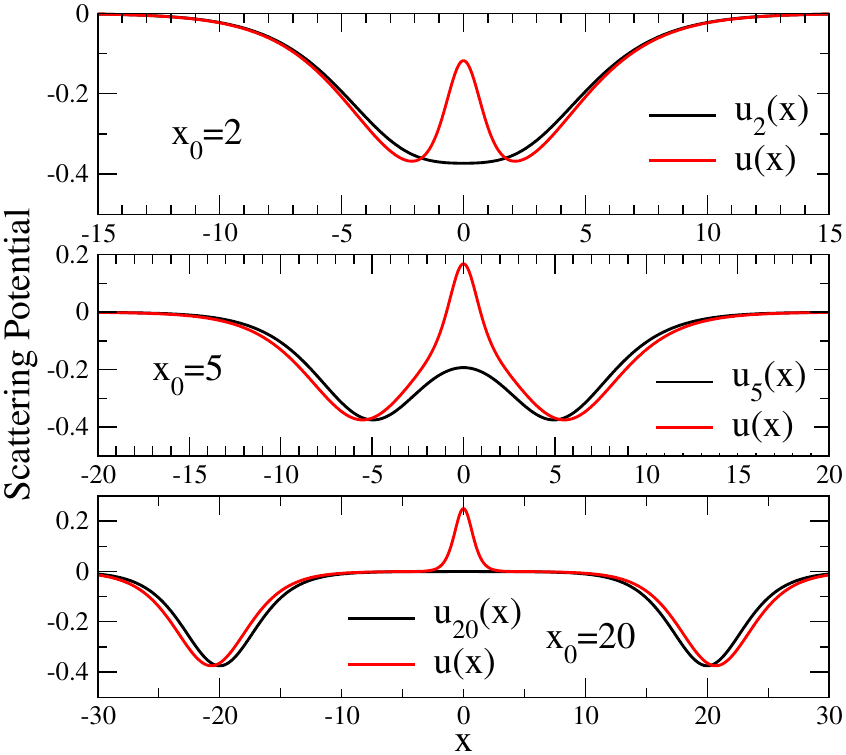,width=5.3cm,height=7.5cm} \hspace{0.3cm}
\epsfig{file=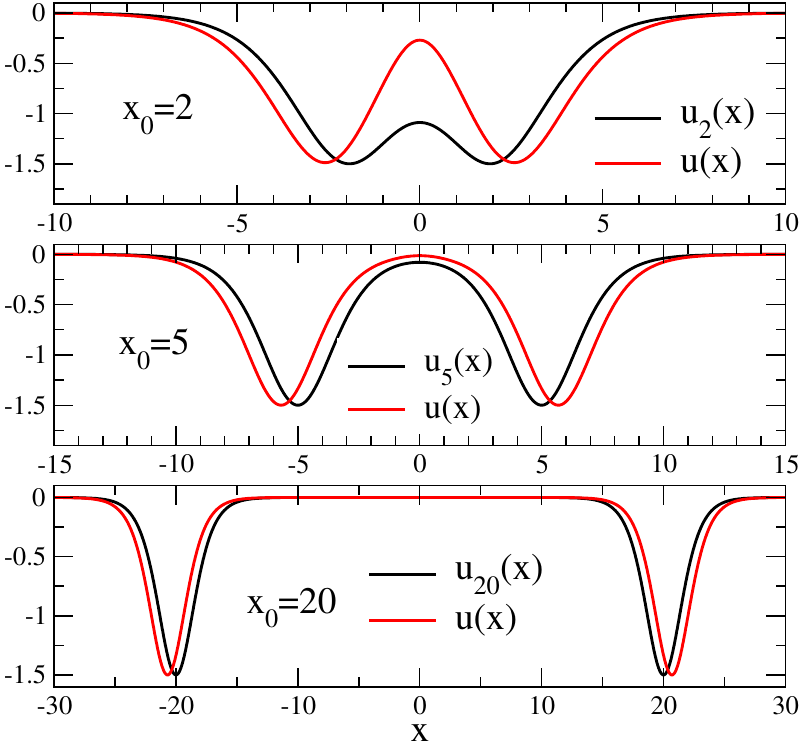,width=5.3cm,height=7.5cm} \hspace{0.3cm}
\epsfig{file=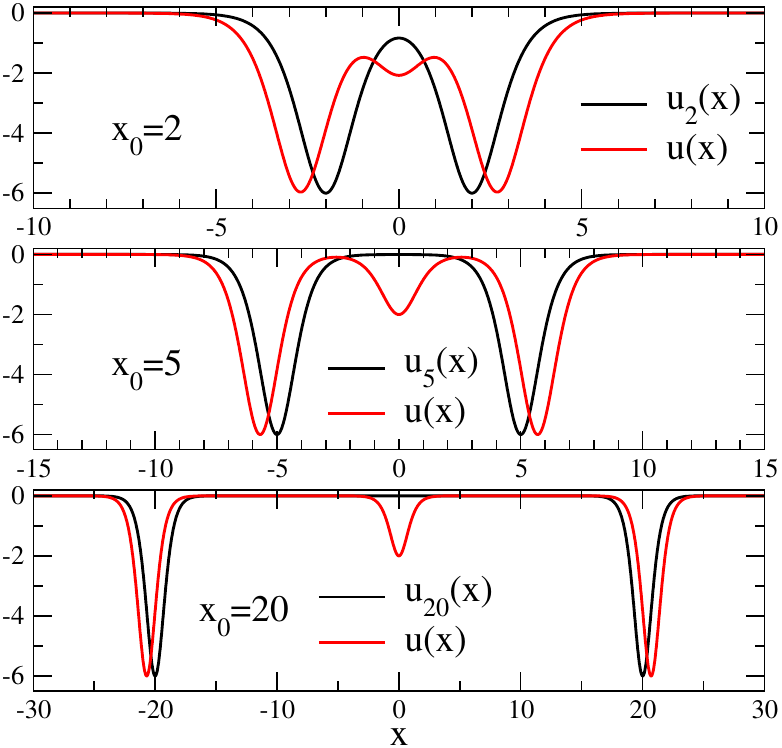,width=5.3cm,height=7.5cm}
}
\caption{\label{fig:compot} The scattering potential $u(x)$ from Eq.~(\ref{eq:scatpot}) compared to 
that of the $\phi^{4}$ configuration $u_{x_0}(x)$ for $x_{0}=2, 5$ and $20$ The latter is defined 
in Eq.~(\ref{eq:KAKpot}). Left to Right: For $j=0.5, 1$ and $2$. Note the different scales along the 
horizontal axis as $x_0$ varies.}
\end{figure}

In the next step, we investigate the behavior of the scattering potential. We do this by considering a small 
perturbation of the BPS solitons $\phi(x,t)=\phi(x)+\eta(x){\rm e}^{-\imu \omega t} $. Substituting this into 
Eq.~\eqref{eq:Lag} and considering linear terms of $\eta$ in the resulting wave equation yields 
\begin{equation}
\left[-\frac{d^2}{dx^2}+V(x)\right]\eta=\omega^{2}\eta\,,
\end{equation}
where $\omega^{2}=m^{2}+k^{2}$ with $k>0$ for the scattering states and $k_{j}=\imu \kappa_{j}$ for the 
bound states. Here $m$ and $k$ are the mass of the fluctuations and the continuous momentum respectively.
Subtracting the analog quantity for $V(x\rightarrow \pm \infty )=m^{2}$ from the wave equation yields 
the scattering wave equation 
\begin{equation}
\left[-\frac{d^2}{dx^2}+u(x)\right]\eta=k^2\eta,
\label{eq:wfeq}\end{equation}
where
\begin{equation}
u(x)=V(x)-m^2=2\left(1+\sigma\right)^{2}
\left[\left(\frac{dW}{d\phi}\right)^2+W\frac{d^2W}{d\phi^{2}}\right]-\sqrt{2}\,\frac{\partial\sigma}{\partial x}
\frac{dW}{d\phi}-m^2
\end{equation}
is the scattering potential. We also identify $m=j$ from $u(x)\,\to\,0$ when $x\,\to\,\infty$. The explicit 
expression for the scattering potential is obtained by substituting the BPS soliton $\phi(x)$, the non-localized 
impurity $\sigma_{j}(x)$ and the super-potential $W(\phi(x))$. This then takes the form
\begin{equation}
u(x)=\frac{j^{2}}{2}\left[\left(3\left[\frac{\cosh^{j}x-a}{\cosh^{j}x+a}\right]^2-1\right)\tanh^2x-2\right] 
-\frac{j}{\cosh^2x}\,\frac{\cosh^{j}x-a}{\cosh^{j}x+a}\,.
\label{eq:scatpot} \end{equation} 
The shape of the scattering potential is shown in Figure \ref{fig:compot}. We observe that for small $j$
the potential has two non-central attractive regions in the vicinity of $\pm x_0$. As we increase $j$
a central structure around the origin emerges which is fully attractive for large $j$. For small and moderate
$j$ it has both attractive and repulsive components.

Since the classical energy does not depend on $x_0$, we expect a zero mode whose wave-function is proportional
to 
\begin{equation}
\eta_0(x)=\frac{\partial \phi(x)}{\partial x_0}
\propto\frac{\partial \phi(x)}{\partial a}=\frac{2\cosh^{j}x}{\left(\cosh^{j}x+a\right)^2}\,.
\label{eq:zeromode}\end{equation}

\section{Scattering potential at large $x_0$}
\label{sec:pot}
We want to investigate the potential for large $a\approx{\rm e}^{jx_0}$ in the three regimes where it 
substantially deviates from zero: (i) $x\approx x_0$, (ii) $x\approx-x_0$ and (iii) $x\approx0$. The first 
two are equivalent by spatial reflection.

For (i) and (ii) we can set $\tanh^2x=1$. To be definite we consider (i) so that
$\cosh^{j}x\approx\frac{1}{2^j}{\rm e}^{jx}={\rm e}^{j(x-\ln2)}$. We then have
\begin{equation}
\frac{\cosh^{j}x-a}{\cosh^{j}x+a}\approx
\frac{{\rm e}^{j(x-\ln2)}-{\rm e}^{jx_0}}{{\rm e}^{j(x-\ln2)}-{\rm e}^{jx_0}}
=\frac{{\rm e}^{j\left(x-x_0-\ln2\right)}-1}{{\rm e}^{j\left(x-x_0-\ln2\right)}+1}
=\tanh\left[\frac{j}{2}\left(x-x_0-\ln2\right)\right]\,.
\label{eq:C1a} \end{equation}
This approximates the potential as
\begin{equation}
u(x)\approx\frac{3}{2}j^2\left\{
\tanh^2\left[\frac{j}{2}\left(x-x_0-\ln2\right)\right]-1\right\}\,,
\label{eq:C1b}\end{equation}
which is a P\"oschl-Teller potential\footnote{These potentials are
$V_l=\frac{l+1}{l}M^2\left[\tanh^2(Mx/l)-1\right]$.\\
The bound state wave-numbers are $t=\sqrt{M^2-E^2}=kM/l$ with $k=0,1,2,\ldots,l$.\\ }
with $M=j$ and $l=2$ centered at $x_0+\ln2$. Its bound state energies have
wave-numbers $t=\sqrt{j^2-\epsilon^2}$ with $t=\frac{j}{2}$ and $t=j$. This approximation also shows
that the zero mode associated with the $x_0$ invariance of the classical energy
turns into the translational zero mode of an ordinary kink.
By a similar analysis we get
\begin{equation}
\phi(x)=-\frac{\cosh^j(x)-a}{\cosh^j(x)+a}\approx-\tanh\left[\frac{j}{2}\left(x-x_0-\ln2\right)\right]
\label{eq:C1c} \end{equation}
for large $a$ and $x\approx x_0$, which again exhibits the antikink structure located at $x_0+\ln2$.
With the kink potential $\frac{\lambda}{4}\left(\phi^2-\frac{M^2}{2\lambda}\right)^2$ this corresponds to
$M=j$ and $\lambda=\frac{j^2}{2}$. Along the same considerations we find a kink structure around $-(x_0+\ln2)$.

\begin{figure}
\centerline{\epsfig{file=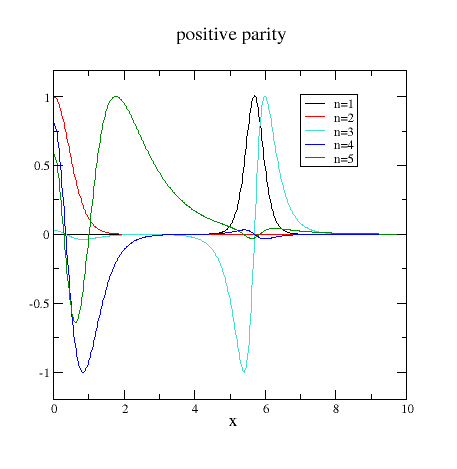,width=7cm,height=5cm}\hspace{1cm}
\epsfig{file=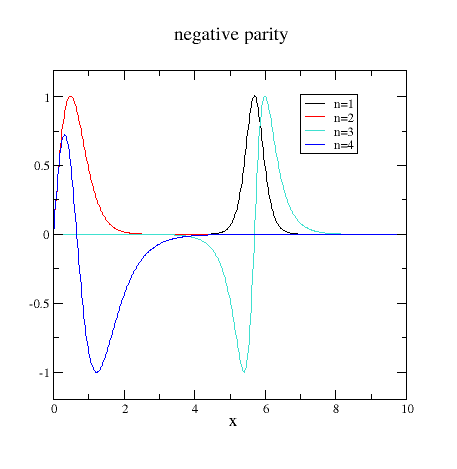,width=7cm,height=5cm}}
\caption{\label{fig:wfct}Bound state wave functions for $x_0=5.0$ and $j=6.0$. The solutions
labeled $n=1$ are the asymptotic ($x_0\,\to\,\infty$) zero modes. We have normalized the wave-functions 
to be in the interval $[-1,1]$.}
\end{figure}

For case (iii) we have $a\gg\cosh^{j}x$ and thus
\begin{equation}
u(x)\approx j(j-1)\left[\tanh^2(x)-1\right]\,.
\label{eq:C2a}\end{equation}
The bound state wave-numbers of this potential are also known\footnote{See, {\it e.g.}
Eqs.~(3.11) and~(3.12) in Ref.~\cite{Gani:2016fqq}.}
\begin{equation}
t=j-n-1
\qquad {\rm with}\qquad
n=0,1,\ldots,\lfloor\, j\rfloor-1\,.
\label{eq:C2b}\end{equation}
The numerical simulation verifies these solutions for large $j$ and $x_0$; including
the zero mode wave-functions being centered at $x_0+\ln2$ as can be seen from the numerical
results shown in figure \ref{fig:wfct}. Since $u(-x)=u(x)$, the bound states have definite parity.

In the process of these numerical simulations we have also verified the existence of the 
zero mode with the wave-function, Eq.~(\ref{eq:zeromode}) for various values of $j$ and $x_0$.

For $j=1$ and large $a$ the central part of the potential (iii) vanishes. Hence we expect that 
in this case the VPE is (twice) that of the kink with $M=1$.

\section{Vacuum polarization energy}
\label{sec:vpe}
The VPE is a measure of the energy change caused by the polarization of the single particle modes, which 
occurs due to the interaction of quantum fluctuations with the background configuration generated by the soliton. 
As already discussed the soliton induces a reflection invariant potential and we can therefore 
write the renormalized boson VPE as
\begin{align}
\Delta E=\frac{1}{2}\sum_{n}^{\rm b.s.}\omega_n
+\int_0^\infty \frac{dk}{2\pi} \sum_{\ell=\pm}
\sqrt{k^2+j^2}\,\left[\frac{d\delta_\ell(k)}{dk}\right]_N
+\sum_{n=1}^N E_{\rm FD}^{(n)}+E_{\rm CT}\,.
\label{eq:almostmaster} \end{align}
The first term sums over the explicit bound states and the momentum integral (with the sum over the
parity channels) collects the continuum contribution. It is the integral over the scattering states
with the relativistic dispersion $\omega=\sqrt{k^2+j^2}$ weighted by the change in the density of 
states expressed by the phase shift according to the Krein formula~\cite{Faulkner:1977zz},
\begin{equation}
\Delta\rho_\ell=\frac{1}{\pi}\frac{d\delta_\ell(k)}{dk}\,.
\label{eq:Krein}
\end{equation}
The subscript $N$ indicates the subtraction of the first $N$ orders of the Born series from the phase
shifts. With $N$ sufficiently large, the momentum integral is finite. These subtractions are added back
in form of equivalent Feynman diagram contributions, $E_{\rm FD}^{(n)}$. Finally, $E_{\rm CT}$ is the
counterterm contribution that implements the renormalization condition(s). Note that 
$\sum_{n=1}^N E_{\rm FD}^{(n)}+E_{\rm CT}$ is also ultraviolet finite. For a boson fluctuation in 
one space dimension the situation is quite simple. There is only one ultraviolet divergent Feynman diagram
with a single insertion of the (Fourier transform at zero momentum of) $u(x)$. It is proportional to 
$\bigintsss_0^\infty dx\,u(x)$ and can thus be canceled completely within the so-called no-tadpole 
renormalization scheme. Hence we write
\begin{align}
\Delta E=\frac{1}{2}\sum_{n}^{\rm b.s.}\omega_n
+\int_0^\infty \frac{dk}{2\pi} \sum_{\ell=\pm}
\sqrt{k^2+j^2}\,\left[\frac{d\delta_\ell(k)}{dk}\right]_1\,.
\label{eq:almostmaster1} \end{align}
The phase shifts are the phases of the Jost functions 
$F_{+}(k)=\lim_{x\to0}\frac{\partial\mathcal{F}(k,x)}{\imu k\partial x}$ and
$F_{-}(k)=\lim_{x\to0}\mathcal{F}(k,x)$ in the two parity channels. Here
$\mathcal{F}(k,x)$ solves the wave-equation, Eq.~(\ref{eq:wfeq}) with the boundary condition
$\lim_{x\to\infty}{\rm e}^{-\imu kx}\mathcal{F}(k,x)=1$. This function is the Jost solution
and is analytic for ${\sf Im}(k)\ge0$ \cite{Newton82,Chadan89}. For real $k$ we also have 
$\mathcal{F}(-k,x)=\mathcal{F}^\ast(k,x)$ so that we express the phase shifts as
an obviously odd function of the momentum:
$\delta_{\ell}(k)=\frac{\imu}{2}\left[\ln F_{\ell}(k) -\ln F_{\ell}(-k)\right]$. In turn we get the 
VPE 
\begin{align}
\Delta E=\frac{1}{2}\sum_{n}^{\rm b.s.}\omega_n
+\frac{\imu}{2}\int_{-\infty}^\infty \frac{dk}{2\pi} 
\sqrt{k^2+j^2}\,\left[\frac{d\ln F_{+}(k)F_{-}(k)}{dk}\right]_1\,.
\label{eq:almostmaster2} \end{align}
This integral can be computed as a contour integral because $F_{\pm}(k)$ is analytic in the upper complex 
momentum plane. As the leading large $|k|$ terms are Born subtracted there is no contribution from the
semi-circle at infinity. The Jost function has simple zeros at the complex bound state momenta 
$k=\imu\sqrt{j^2-\omega_n^2}$ so that the singularities from the logarithmic derivative cancel
$\frac{1}{2}\sum_{n}^{\rm b.s.}\omega_n$. The only contribution stems from bypassing the branch cut induced 
by the relativistic dispersion relation along the imaginary axis $t\in\left[j,\infty\right]$ where
$k=\imu t$. To efficiently compute that final integral we factorize 
$\mathcal{F}(\imu t,x)={\rm e}^{-tx}g(t,x)$ and solve the differential equation
\begin{equation}
\frac{\partial^2 g(t,x)}{\partial x^2}=2t\frac{\partial g(t,x)}{\partial x}+u(x)g(t,x)\,,
\label{eq:DEQg}\end{equation}
with boundary condition $g(t,\infty)=1$. After a final integration by parts we get
\begin{equation}
\Delta E=\int_{j}^{\infty}\frac{dt}{2\pi}\frac{t}{\sqrt{t^{2}-j^{2}}}
\left\{\ln\left[g(t,0)\left(g(t,0)-\frac{1}{t}\frac{\partial g(t,x)}{\partial x}\Big|_{x=0}\right)\right]
-\frac{1}{t}\int_0^\infty dx\, u(x)\right\}\,.
\label{eq:vpe} \end{equation}
We have made the Born subtraction explicit. The coefficient of that spatial integral can be most straightforwardly 
derived by expanding $g(t,x)=1+g^{(1)}(t,x)+\mathcal{O}(u^2)$ and integrating the corresponding differential
equation
$$
\frac{\partial^2 g^{(1)}(t,x)}{\partial x^2}=2t\frac{\partial g^{(1)}(t,x)}{\partial x}+u(x)\,,
$$
with the boundary condition $g^{(1)}(t,\infty)=0$ along the positive half axis in coordinate space.

\begin{table}
\centerline{
\begin{tabular}{c| c c c c c c c c c c c c  }
\hline \hline 
$x_{0}$ &  1.5  & 4.0  &  8.0  & 12.0  &  16.0  & 20.0  & 24.0 & 28.0 & 32.0 & 36.0 & 40.0 & 44.0 \cr \hline 
$\Delta E$ & -0.06455 & -0.06992  & -0.09204  & -0.11242  & -0.12575 & -0.13312  & -0.13681 & -0.13856 
& -0.13937 & -0.13974  & -0.13990 & -0.13995 
\end{tabular}}
\caption{\label{t0}VPE as function of $x_{0}$ for $j=0.2$.}
\end{table}

\section{Numerical Results for the VPE}
\label{sec:results}
\begin{table}
\centerline{
\begin{tabular}{c| c c c c c c c c c c  }
\hline \hline 
$x_{0}$ &  1.5  & 2.0  &  3.0  & 4.0  &  5.0  & 6.0  & 7.0 & 8.0 & 9.0 & 10.0 \cr \hline 
$\Delta E$ & -0.17786 & -0.19343 & -0.22755 & -0.25882 & -0.28383 & -0.30221 & -0.31493 
& -0.32336 & -0.32878 & -0.33219 \cr \hline \hline 
$x_{0}$ &  11.0  & 12.0  &  13.0  & 14.0  &  15.0  & 16.0  & 17.0 & 18.0 & 19.0 & 20.0 \cr \hline 
$\Delta E$ & -0.33431 & -0.33562 & -0.33642 & -0.33691  & -0.33721  & -0.33739 &  -0.33749  
& -0.33754 &  -0.33756 & -0.33758 
\end{tabular}}
\caption{\label{t1}Behavior of the VPE as function of $x_{0}$ for $j=0.5$.}
\end{table}
In this section, we report the results from our numerical simulations of the one-loop VPE discussed above. 
We solve Eq. \eqref{eq:DEQg} using the fourth order Runge Kutta algorithm with an adaptive step size 
control for various values of $j$ and $x_0$. That is, we compute the quantum correction to the energy as a 
function of the variational parameter, $x_{0}$ with respect to which the classical energy is degenerate. 

As indicated earlier, the BPS solution behaves like a superposition of the kink-antikink configuration in 
the $\phi^{4}$ model approximated by 
\begin{equation}
\phi_{R}(x)=\tanh\left[\frac{j}{2}\left(x+R\right)\right]-\tanh\left[\frac{j}{2}\left(x-R\right)\right]-1 
\label{eq:ansatz_config} \end{equation}
where $R$ indicates the position of the kink. This configuration gives rise to the scattering potential 
\begin{equation}
u_{R}(x) = \frac{3j^2}{2}\left[\phi_{R}^{2}(x)-1\right].
\label{eq:KAKpot}\end{equation}
\begin{figure}
\centerline{
\epsfig{file=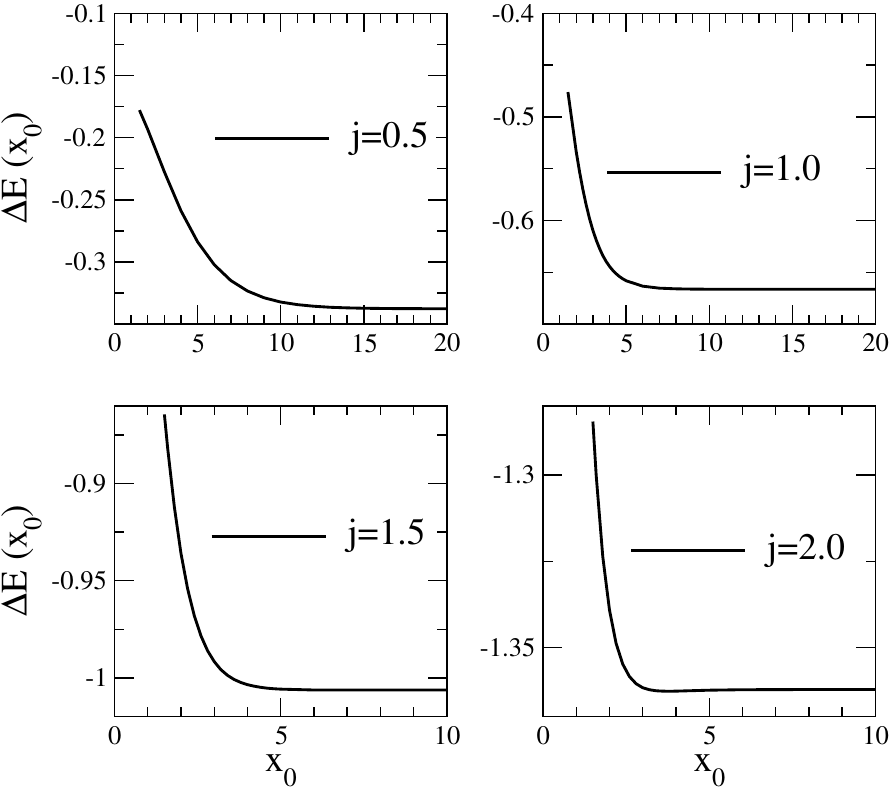,width=8.5cm,height=7.5cm} \hspace{0.2cm}
\epsfig{file=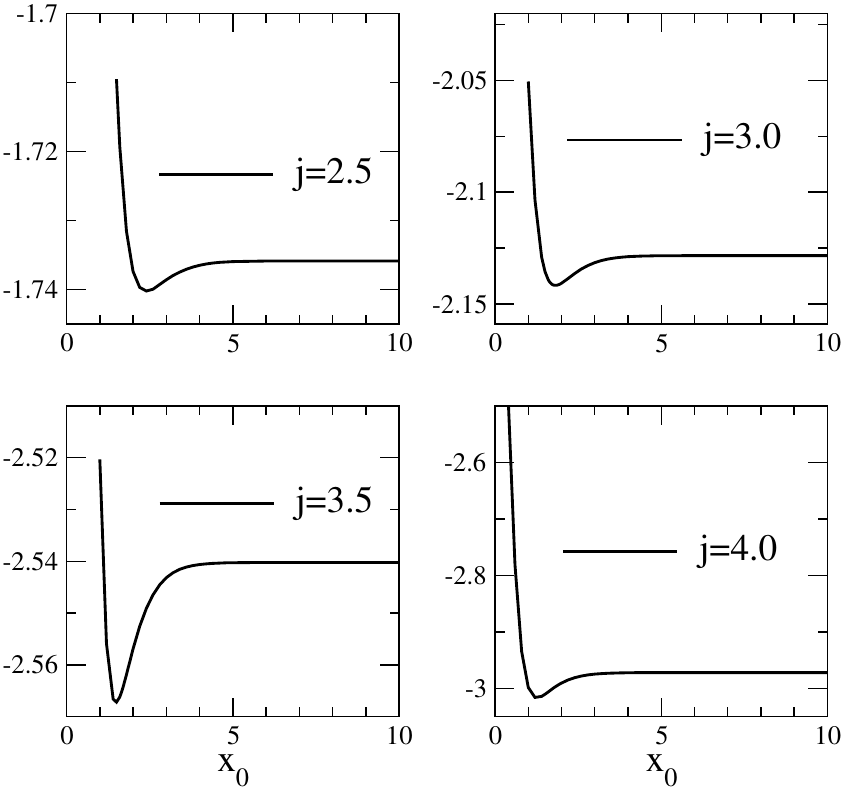,width=8.5cm,height=7.5cm}}
\caption{\label{fig:vacuumenergies} The vacuum polarization energy as a function of $x_{0}$ for
various values of $j$. Note the different horizontal scales for small $j$.} 
\end{figure}
In Fig. \ref{fig:compot} we also compare this to the BPS scattering potential. We see that $j=1$ best approximates 
the kink-antikink configuration of the $\phi^4$ model for $x_0\gg1 $. This confirms the approximation, Eq.~(\ref{eq:C2a}) 
according to which the central structure vanishes for this value of $j$ when $x_0$ is large. The only difference is that 
the position of the potential minimum is somewhat larger than $R$; of course this is just the additional $\ln2$ identified
in section \ref{sec:pot}. For $j\,{\scriptstyle \gtrsim}\,5$ the non-central structures of $u(x)$ almost 
exactly match $u_R(x)$, with the same small deviation of the minimum position from~$R$. 

We display the results for $\Delta E$ as a function of $x_{0}$ from our numerical simulations for
other impurity values in tables \ref{t0}-\ref{t4} each referring to different values of $j$. We 
observe that  $\Delta E$ decreases when increasing $x_{0}$ from a small finite value to infinity. For 
small $j$ it does so monotonously but develops a tiny local minimum at a moderate $x_{0}$ when $j$ 
increases.  This is due to the attractive central structure of $u(x)$. The position of the local 
minimum moves to smaller values of $x_0$ as $j$ increases. This is also obvious from the graphs 
in Fig.\ref{fig:vacuumenergies}. Hence we conclude that for a significantly strong impurity
the quantum corrections resolve the classical degeneracy while for weak impurities $x_0\to\infty$
is favored which causes the soliton to be unstable. From Fig. \ref{fig:vacuumenergies} and table 
\ref{t3} we immediately recognize an energy minimum for $j=2$. For $j=1.5$ the existence of such a very 
shallow minimum is not obvious from the figure. However, the data $\Delta E(7.0)-\Delta E(15.0)=-0.0004$ indeed 
indicate the existence of a minimum already for $j=1.5$. Even though this small difference is of the order of 
the numerical accuracy we conjecture that minima emerge for all $j>1$ since for this value and large 
enough $x_0$ the central structure of $u(x)$ turns from repulsive to attractive at $j=1$; as discussed in 
Sect. \ref{sec:pot}. Eventually the VPE saturates when $x_0$ is large enough because the various structures 
in $u(x)$ separate without changing their shapes as $x_0$ grows further. For weak impurities the asymptotic 
value is (approximately) reached only for very large $x_0$ but the saturation position decreases as $j$ increases. 
Again, this can be understood from the discussion in Sect. \ref{sec:pot}: the larger $j$, the better the 
approximation $\cosh^j(x-x_0)\approx {\rm e}^{j(x-x_0-\ln2)}$.

With increasing impurity strength the local minimum in $\Delta E$ moves to smaller $x_0$ values and
the $x_0$ behavior of the soliton profile shown in figure \ref{fig:profile} suggests that the soliton 
eventually loses its kink-antikink shape.

The case $j=1$ is particularly interesting because we have conjectured that the central structure
becomes irrelevant as $x_0$ increases and that the potential in the vicinity of $\pm x_0$ is 
of P\"oschl-Teller type with $l=2$. That is exactly the background potential induced by the
kink in a $\phi^4$ model. And indeed we see from table \ref{t2} that $\Delta E$ approaches 
$-0.66623$ asymptotically which is twice the Dashen-Hasslacher-Neveu \cite{Dashen:1974ci} value 
$\textstyle\left(\frac{1}{4\sqrt{3}}-\frac{3}{2\pi}\right)\approx-0.333127$ for the VPE of 
the $\phi^4$ kink soliton for unit mass parameter.
\begin{table}
\centerline{
\begin{tabular}{c| c c c c c c c c c c   }
\hline \hline 
$x_{0}$ &  1.5  & 2.0  &  3.0  & 4.0  &  5.0  & 6.0  & 7.0 & 8.0 & 9.0 & 10.0 \cr \hline 
$\Delta E$ & -0.475999 & -0.53358  & -0.60935 & -0.64415 &  -0.65805 & -0.66328  & -0.66523 
& -0.66599  & -0.66617 & -0.66622  \cr \hline \hline 
$x_{0}$ &  11.0  & 12.0  &  13.0  & 14.0  &  15.0  & 16.0  & 17.0 & 18.0 & 19.0 & 20.0  \cr \hline 
$\Delta E$ & -0.66623 & -0.66623 & -0.66623 & -0.66623 & -0.66623 & -0.66623 & -0.66623 
& -0.66623 & -0.66623 & -0.66623 
\end{tabular}}
\caption{\label{t2}VPE as function of $x_{0}$ for $j=1.0$, whose asymptotic value is twice the VPE value 
of a single kink of the $\phi^{4}$ model. }
\end{table}
Though supported by the analytical considerations in Sect.\ref{sec:pot} in the limit of large $x_0$, this 
agreement occurs already for moderate $x_0$. For example, for $x_0=5$ the difference in the VPEs is 
just about 1\%.

\begin{table}
\centerline{
\begin{tabular}{c| c c c c c c c c c    }
\hline \hline 
$x_{0}$ &  1.5  & 2.0  &  3.0  & 4.0  &  5.0  & 6.0  & 7.0 & 8.0 & 9.0   \cr \hline 
$\Delta E$ & -1.28454 & -1.33903  & -1.36159  & -1.36259  &  -1.36230  & -1.36217 
& -1.36214 & -1.36213 & -1.36213   \cr \hline \hline 
$x_{0}$ &  10.0  & 11.0  &  12.0  & 13.0  &  14.0  & 15.0  & 16.0 & 17.0 & 18.0   \cr \hline 
$\Delta E$ & -1.36213 & -1.36213  & -1.36213  & -1.36213 & -1.36213 & -1.36213 & -1.36213 &  -1.36213 &  -1.36213
\end{tabular}}
\caption{\label{t3}VPE as function of $x_{0}$ for $j=2.0$.}
\end{table}
\begin{table}
\centerline{
\begin{tabular}{c| c c c c c c c c c    }
\hline \hline 
$x_{0}$ &  1.5  & 2.0  &  3.0  & 4.0  &  5.0  & 6.0  & 7.0 & 8.0 & 9.0  \cr \hline 
$\Delta E$ & -1.70952 & -1.73737 & -1.73857  & -1.73648  &  -1.73594 & -1.73586  
& -1.73585  & -1.73585 & -1.73585    \cr \hline \hline 
$x_{0}$ &  10.0  & 11.0  &  12.0  & 13.0  &  14.0  & 15.0  & 16.0 & 17.0 & 18.0   \cr \hline 
$\Delta E$ & -1.73585  & -1.73585  & -1.73585  & -1.73585  & -1.73585  & -1.73585  
& -1.73585 &  -1.73585  &  -1.73585  
\end{tabular}}
\caption{\label{t4}VPE as function of $x_{0}$ for $j=2.5$.}
\end{table}

\section{Quantum corrections to the kink-antikink potential}
\label{sec:KAK}

We identify the scattering potential associated with the central region by defining
\begin{equation}
\widetilde{u}(x)=u(x)\begin{cases} 1\,, & \quad |x|\le \frac{x_0}{2} \cr
{\rm e}^{-(|x|-x_0/2)^2/w^2}\,, & \quad |x|> \frac{x_0}{2}\end{cases}
\label{eq:potcenter}\end{equation}
and compute the corresponding VPE, $\widetilde{\Delta E}$. For large enough $x_0$ this
is not sensitive to the width parameter $w$ as long as it still small compared to $x_0$. 
The quantum correction to the kink-antikink potential is the difference
\begin{equation}
\Delta E_{K\overline{K}}=\Delta E-\widetilde{\Delta E}\,.
\label{eq:EKK}\end{equation}
A direct identification of this correction would lead to imaginary frequencies (equivalently, zeros
of the Jost function for $t>j$) for the would-be zero mode(s) because the static configuration from 
Eq.~(\ref{eq:ansatz_config}) is not a solution to the kink wave-equation. In a sense we can interpret the 
central structure as the source needed to keep the $\phi^4$ (anti)kink in place \cite{Bashinsky:1999vg}.

\begin{figure}
\centerline{
\epsfig{file=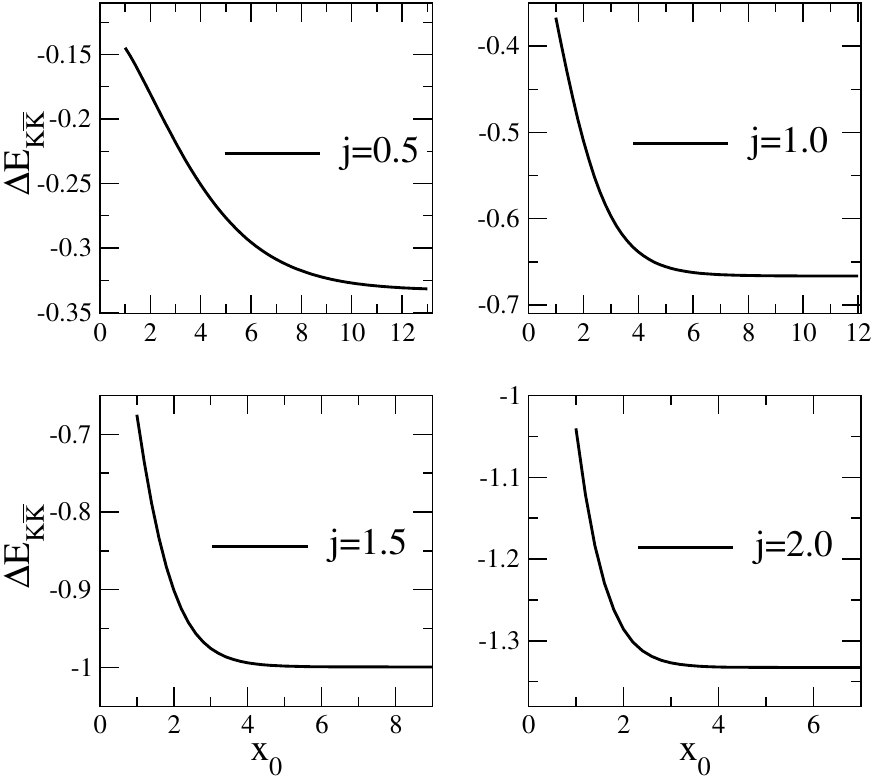,width=8.5cm,height=7.5cm} \hspace{0.2cm}
\epsfig{file=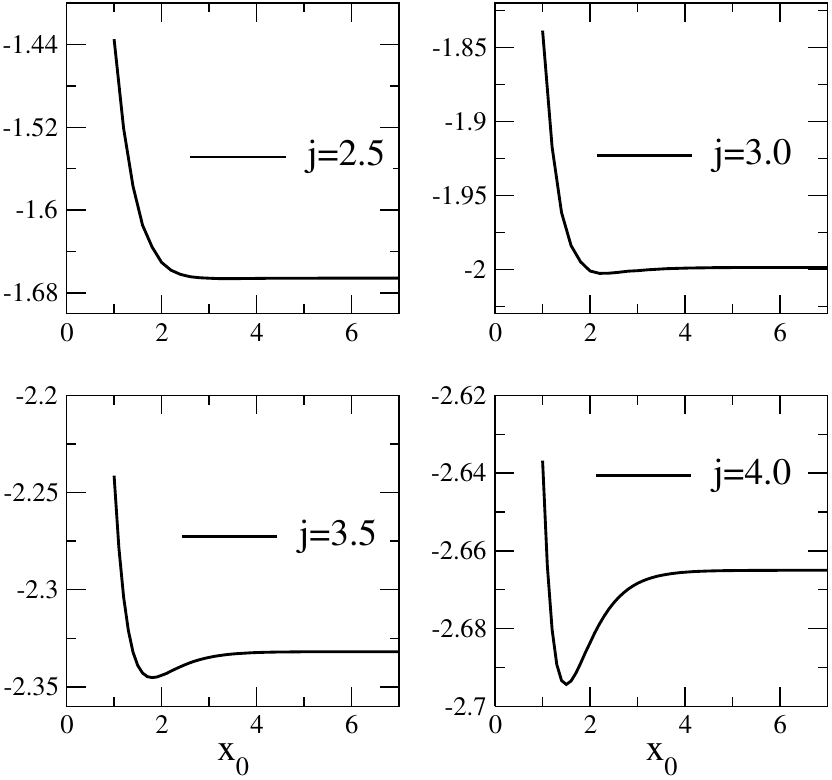,width=8.5cm,height=7.5cm}}
\caption{\label{fig:VPEKKbar} The quantum correction to the kink-antikink potential, Eq.~(\ref{eq:EKK})
as a function of $x_{0}$ and various values of $j$.}
\end{figure}

The numerical results for $\Delta E_{K\overline{K}}$ as functions of $x_0$ for different strengths $j$ of
the impurity are shown in tables \ref{tab:KK1}-\ref{tab:KK3} and figure \ref{fig:VPEKKbar}. In all cases we 
reproduce the expected asymptotic value 
$\Delta E_{K\overline{K}}\,\to\,j\left(\frac{1}{2\sqrt{3}}-\frac{3}{\pi}\right)$ for $x_0\,\to\,\infty$ 
suggested via the P\"oschl-Teller potential in Eq.~(\ref{eq:C1b}).

\begin{table}
\centerline{
\begin{tabular}{c| c c c c c c c }
\hline \hline
$x_{0}$ &  1.0  & 3.0  &  5.0  & 7.0  &  9.0  & 11.0  & 13.0 \cr \hline
$\Delta E_{K\overline{K}}$ & -0.1448 & -0.2182 & -0.2766  & -0.3088 &  -0.3235 & -0.3295 &-0.3317
\end{tabular}}
\caption{\label{tab:KK1}The quantum correction to the kink-antikink potential, Eq.~(\ref{eq:EKK}) 
as a function of $x_{0}$ for $j=0.5$ and $w=2.0$.}
\end{table}
\begin{table}
\centerline{
\begin{tabular}{c| c c c c c c c }
\hline \hline
$x_{0}$ &  1.0  & 2.0  &  3.0  & 4.0  &  5.0  & 6.0  & 7.0 \cr \hline
$\Delta E_{K\overline{K}}$ & -1.4350 & -1.6505 & -1.6657  & -1.6659 & -1.6657 & -1.6656 &-1.6656
\end{tabular}}
\caption{\label{tab:KK2}Same as Tab. \ref{tab:KK1} for for $j=2.5$.}
\end{table}
\begin{table}
\centerline{
\begin{tabular}{c| c c c c c c c }
\hline \hline
$x_{0}$ &  1.0  & 2.0  &  3.0  & 4.0  &  5.0  & 6.0  & 7.0 \cr \hline
$\Delta E_{K\overline{K}}$ & -2.6254 & -2.6836 & -2.6684  & -2.6655 & -2.6651 & -2.6650 &-2.6650
\end{tabular}}
\caption{\label{tab:KK3}Same as Tab. \ref{tab:KK1} for for $j=4.0$.}
\end{table}
As in the case of $\Delta E$, $\Delta E_{K\overline{K}}$ is purely repulsive for small and moderate $j$.
However, for $j\,{\scriptstyle \gtrsim}\,2$ a local minimum emerges. The $x_0$ position of this minium approaches 
the center as $j$ increases. We therefore conclude that the attraction for larger $j$ is not only a property of
the central structure of the fluctuation potential but also signals an attractive quantum contribution to the
kink-antikink interaction. The energies shown in figure \ref{fig:VPEKKbar} suggest that the interaction becomes 
arbitrarily strongly repulsive as $x_0$ approaches zero. Thus would eventually prevent the kink-antikink system from 
collapsing (as it may in the pure $\phi^4$ model). Unfortunately, extracting this interaction via Eq.~(\ref{eq:EKK}) 
becomes an invalid procedure in that interesting limit.

\section{Conclusion}
\label{sec:conclude}
We have computed the leading (one-loop) quantum correction to soliton energies in an BPS-impurity model in one 
time and one space dimensions.  Our motivation for this study was two-fold. First, with the model parameters 
set to the BPS-case, the classical energy is degenerate with respect to a continuous real parameter. Usually 
one trusts the computation of the leading quantum correction only when it is (significantly) smaller than
the classical counterparts; otherwise it is very likely that even higher order corrections are equally
important and should not be omitted. (See {\it e.g.} Ref. \cite{Evslin:2021vgk} for estimates beyond
one loop.) However, when comparing classically degenerate configurations, the 
classical energy is irrelevant and the quantum corrections are decisive for selecting the favorable 
configuration. Second, the above mentioned continuous parameter can be associated with the separation of a 
kink-antikink pair in the renowned $\phi^4$ kink model, at least for moderate and large values. This allows 
to investigate the quantum corrections to the kink-antikink interaction which, due the emergence of unstable 
fluctuation modes, is unfortunately not directly possible in the $\phi^4$ kink model.

In this model the soliton solution generates a potential for the quantum fluctuations with three structures. 
We have identified these structure from analytic considerations and also by exploring the bound state structure. 
Two of these structures can be associated with the kink and the antikink at the respective positions while the 
impurity induces a central structure in-between. For small strengths of the impurity the central structure is 
repulsive but gets more and more attractive as this strength is increased. As consequences, for weak impurities 
the quantum fluctuations destabilize the soliton in the sense that it is energetically favorable to pull the kink 
and the antikink components infinitely far apart. This means that the soliton occupies a secondary vacuum in an 
ever increasing region of space. Hence this instability is conceptually similar to the one previously observed in the 
Shifman-Voloshin model. As the strength increases the potential extracted from the quantum fluctuations develops a 
minimum which determines the favorable value of the parameter with respect to which the classical energy is degenerate. 
The more the impurity strength is increased the more the soliton loses its kink-antikink shape.

When we remove the central structure from the potential, we can get some insight into the quantum corrections
to the kink-antikink interaction in the $\phi^4$ model. Unfortunately, the corresponding results depend on the
impurity strength so that we cannot make a general statement. However, we find that these corrections are
mostly repulsive, only for large strengths a moderate attraction occurs.

We note that the choice for the impurity is somewhat arbitrary. The choice considered in Ref.~\cite{Evslin:2022xmp} 
generates an antikink. Similar to the present case, that classical energy does not dependent on the distance
between the centers of the antikink and the impurity. Using the technique\footnote{That technique relates the creation 
and annihilation operators for the quantum fluctuations with and without a soliton background. For the standard 
kink Ref.~\cite{Evslin:2019xte} reproduced the well established historic result \cite{Dashen:1974ci} and thus agrees 
with the spectral method \cite{Graham:2009zz}.} of Ref.~\cite{Evslin:2019xte} those authors hence computed the quantum 
correction to the energy as a function of that distance but did not observe a local minimum. It would be interesting 
to verify that result using the spectral method approach. Eventually the comparison of the VPEs for various impurities 
can further disentangle the effects stemming from the soliton on one side and the impurity on the other.

\acknowledgments
H.\@ W.\@ is supported in part by the National Research Foundation of
South Africa (NRF) by grant~109497. The authors thank C. Halcrow for bringing
Ref.\@\cite{Evslin:2022xmp} to their attention.

\end{document}